\documentclass[preprint,review]{elsarticle}

\usepackage{graphicx}
\usepackage{amssymb}
\usepackage{natbib}
\usepackage{lineno}
\usepackage{multirow}
\usepackage{hhline}
\usepackage{subfigure}
\usepackage{rotating}

\usepackage{fancyhdr}
\fancypagestyle{pprintTitle}{%
  \lhead{}\lfoot{
    DOI: http://dx.doi.org/10.1016/j.nima.2017.03.064 \\
 \ \\
    \copyright \ 2017. This manuscript version is made available under the CC-BY-NC-ND 4.0 license http://creativecommons.org/licenses/by-nc-nd/4.0/}\cfoot{}\rfoot{}
  
}
\journal{NIMA}

\begin{document}
\newcommand{\Angst}{$\mathring{\mathrm{A}}$}

\begin{frontmatter}

\title{A Polyethylene-B$_{4}$C based concrete for enhanced neutron shielding at neutron research facilities}

\author[a,b]{D. D. DiJulio\corref{cor1}}
\ead{Douglas.DiJulio@esss.se}

\author[a,c]{C. P. Cooper-Jensen}

\author[a,b]{H. Perrey}

\author[a,b]{K. Fissum}

\author[b]{E. Rofors}

\author[a,b]{J. Scherzinger}

\author[a,c]{P. M. Bentley}

\cortext[cor1]{Corresponding Author}

\address[a]{European Spallation Source ERIC, P.O. Box 176, SE-221 00 Lund, Sweden}
\address[b]{Division of Nuclear Physics, Lund University, SE-221 00 Lund, Sweden} 
\address[c]{Department of Physics and Astronomy, Uppsala University, Sweden}

\begin{abstract}
We present the development of a specialized concrete for neutron shielding at neutron research facilities, based on the addition of
hydrogen atoms in the form of polyethylene and also B$_{4}$C for enhancing the neutron
capture properties of the concrete. We show information on the mechanical properties of the concrete and the neutronics,
in particular it's relevance to modern spallation neutron sources, such as the European Spallation Source (ESS), currently under construction
in Lund, Sweden. The new concrete exhibits a 15\% lower mass density, a compressible strength
of 50\% relative to a standard concrete and a significant increase in performance of shielding against MeV neutrons and lower energies. 
The concrete could find application at the ESS in for example common shielding components, individual beamline shielding and instrument caves.
Initial neutronic tests of the concrete, carried out at Lund University, have also verified the performance in the MeV neutron energy range
and the results are presented.
\end{abstract}

%
%


\hyphenation{}

\end{frontmatter}


\section{Introduction}
Neutron research facilities offer scientists the opportunity to use thermal and cold neutrons for scientific
investigations in both fundamental and applied sciences. At spallation neutron sources, such as the European Spallation Source (ESS) \cite{ESS}, currently
under construction in Lund, Sweden, the creation of these cold and thermal neutrons involves the bombardment of a heavy metal target with
an intense and high-energy proton beam. The neutrons that escape the target can undergo scattering in moderators placed in
the vicinity of the target and are ultimately guided towards the sample position of a neutron instrument, which can be up to
around a couple of hundred meters away from the target. However, the neutrons that are not slowed down, in addition to other high-energy particles, can travel deep
into the shielding, which surrounds the target and moderators, and induce the creation of secondary particles that could also
reach the location of a neutron instrument. These unwanted particles are time-correlated with the proton beam
pulse and may lead to backgrounds which make measuring weak signals impossible or challenging \cite{Cherkashyna2015japan}.
Therefore, neutron shielding not only plays an important role for radiation safety purposes, but also for
minimizing the unwanted background noise in neutron instruments.

The bulk shielding at spallation neutron sources typically consists of steel and concrete. The choice of these materials is due to a number of different reasons. The attenuation of high-energy neutrons, above 10s of MeVs, is primarily driven by the material density, and therefore materials which are based on metals perform effectively. Below this energy range, hydrogen containing materials are well suited for the slowing down of neutrons. Compared to iron, which requires 410 collisions to slow a neutron down from 2 MeV to 1 eV, only 15 collisions are required for hydrogen \cite{Bauer2001}. Thus hydrogen containing materials such as concrete can be used to effectively reduce the energy of incident neutrons, especially around a few MeV and lower, to a regime where the absorption cross-section is larger for most materials. Furthermore, in this energy range, metal-based materials exhibit a significant number of pronounced minima in their cross-sections, called windows \cite{Moreh2006}. Any neutron which is scattered into one of these windows will have a long mean-free path in the material and can travel a long distance. The addition of concrete shielding thus also serves the role of blocking the windows in the metal-based shielding. The concrete shielding can also often contain boron for instance, as in the outer layers of the shielding monolith at the Swiss Spallation Neutron Source (SINQ \cite{SINQ,Wagner2001}) at the Paul Scherrer Institute, Villigen, Switzerland \cite{PSI}, in order to absorb the neutrons which have slowed down to thermal energies. The final selection of the shielding materials is then a trade-off between these physics properties
and for example the cost of the materials, floor loading and operational requirements, just to name a few.         

Both experimental and theoretical evidence for the features described above have been observed in for example earlier studies \cite{Nunomiya2001,Nakao2004}
at the ISIS spallation neutron source \cite{ISIS} at the Rutherford Appleton Laboratory \cite{RAL}.
An overview of the bulk shielding materials at this spallation source, consisting of iron and concrete, can be found within the Refs. \cite{Nunomiya2001,Nakao2004}
where simulations and measurements revealed that the penetration energy spectra of neutrons through the bulk shielding had a broad component which peaks between 10$^{-4}$ and 10 MeV, corresponding to the slowing down of neutrons in the concrete and iron shielding. Additionally, a recent survey carried out at the Spallation Neutron Source (SNS), Oakridge, Tennessee, USA \cite{SNS} also suggested that the dominant component of neutron leakage through the bulk shielding around the target was likely related to neutrons on the order of a few MeVs \cite{DiJulio2016sns}. Therefore, continued research and development into new shielding solutions could be considered of great interest to both current modern and future spallation neutron sources. 

For the above mentioned reasons, we investigated and developed a new specialized concrete for enhanced neutron shielding. The new design was based on increasing the hydrogen content in concrete with the final aim of an improved minimization of neutrons in the MeV and lower energy ranges and with the goal of increasing neutron instrument performance at spallation neutron sources. In the following, we present a description of the concrete, including it's neutronic and mechanical properties, followed by an experimental verification of the concrete in the MeV neutron range.

\section{Design of the Concrete}
\subsection{Performance at a spallation neutron source}

The main motivation behind the concrete was to increase the concentration of hydrogen through the addition of polyethylene (PE), in addition to boron in the form of B$_{4}$C. Early studies of adding hydrogen to concrete, in the form of water, found that a 7 wt\% water content was sufficient for the slowing down of intermediate-energy neutrons to the thermal energy range at a nuclear reactor \cite{Oakridge1958}. It was also concluded that a greater water content would only improve the neutron attenuation properties of the concrete. The idea of adding boron to concrete is also well known. Early efforts were aimed at reducing the dose behind a neutron shield due to neutron capture photons through the addition of boron and a detailed early overview of work on specialized neutron shielding concretes can be found in \cite{Oakridge1970}. In some more recent studies, Ref. \cite{Kratochvil2014} investigated how the additions of different boron compounds (up to 2\% of the cement mass) affected the setting of Portland cement and Ref. \cite{Palomino2008} reported on the neutronic benefits
of a plastic neutron shielding containing B$_{4}$C for thermal and epithermal neutrons. A replacement for heavy concrete at neutron research facilities was proposed in Ref. \cite{Calzada2011}, where the authors presented
a re-usable steel resin, paraffin/PE and boron mixture for neutron shielding. Of specific interest to this study, the addition of both PE and B$_{4}$C to concrete was investigated in detail previously in Ref. \cite{Park2014} however no results on the actual construction or neutronic testing of the proposed PE-B$_{4}$C-based concrete were presented.

To investigate the potential for using such a concrete at a spallation neutron source, we carried out simulations of the performance of the concrete over the entire relevant energy range. For these purposes, we used an incident energy spectrum from a Geant4 \cite{GEANT4,Geant4Ref1} model based on the technical-design-report concept of the ESS \cite{DiJulio2016ess}. This spectrum represents the total number of neutrons over four 60$^o$ beamline extraction areas, at 2 m from the moderator position. This spectrum was then used as input to simulations and incident on various concrete blocks, of different compositions, and with a thickness of 25 cm. Each block was $\pm$25 m in dimensions in the lateral directions and was followed by a detector of the same lateral dimensions, placed directly behind the position of a block. The simulation setup is shown in Fig. 1.

While the concrete is not planned to be used as a beamstop, these simulations illustrate the effects of modifying the composition of concrete over a wide energy range. The calculations were also carried out using Geant4 and the composition of the two concretes which were investigated are given in Table 1, which is based on the mixing recipe presented in the next section. The physics list QGSP$\_{}$BERT$\_{}$HP \cite{Geant4physics} was used for the simulations. Here, QGS refers to the quark-gluon string model, P to pre-compound, BERT to the Bertini cascade model \cite{Bert63,Bert69}, and HP to the high-precision neutron package \cite{Geant4physics}.

The results of the simulations are shown in Fig. 2. The green curve indicates the incident ESS beam and the other curves show the energy of the spectra of the neutrons which have penetrated through the indicated material compositions and thicknesses. These results include all neutrons which arrived at the detector position, including the transmitted and scattered components. The standard reference concrete is indicated by the red line, where one can see the thermal peak at low energies. The purple curve shows the result of adding
10\% by weight of PE to the reference concrete, where it can be seen that while the thermal peak still remains, the region between 1 eV and 1 MeV is greatly reduced. If only B$_{4}$C is added to the reference concrete (blue curve), the thermal peak is removed but little effect is seen between 1 eV and 1 MeV. The black curve represents the addition of both the PE and B$_{4}$C (refered to as PE-B4C-concrete), where the combined effect of these
two components is clearly seen in the new concrete. Our calculations also showed that in order to reach the same level of performance for 1 eV to 1 MeV neutrons as PE-B4C-concrete, 60\% more reference concrete or B$_{4}$C-only concrete would be needed. This illustrates one of the major benefits of such a concrete, that it would be both lighter and less material would be needed in order to match the perfromance of thicker blocks of more standard types of concretes.

We also investigated the simulated prompt-photon production due to neutron interactions in the different concrete material compositions, and the results are shown in Fig. 3. The simulation conditions were the same as described above. The addition of the PE to the concrete had little effect on the photon production and thus the figure only presents results for the reference concrete and the PE-B4C-concrete. The addition of the B$_{4}$C resulted in approximately a third of the photon production compared to the standard concrete. Above 0.5 MeV, this reduction is closer to an order of magnitude. These results indicate that such a concrete can also be used to lower the radiation dose due to photons outside of a thick shield, as discussed earlier.

It should be mentioned that these calculations assume that the PE and B$_4$C are homogeneously distributed throughout the concrete block. The effects of inhomogeneities and grain sizes of the additives, for example as seen in previous studies of Boral are not considered \cite{Oakridge1960,Sweden1960}.

\subsection{Construction of the concrete}

Inspired by the above presented simulations, we had a concrete mixture produced based on the addition of PE and B$_4$C. The recipe for the concrete is shown in Table 2, for both the reference concrete and the new concrete.
For each ingredient, both the weight and approximate volume percent is indicated. The volume is approximate due to small uncertainties in the mass densities of some of the ingredients. The mixing of the concrete was performed by the Danish Technological Institute (DTI) \cite{DTI,DTIreport} and it was found that a 50-50 mix by weight of 2.5 mm and 5.0 mm PE beads provided a homogeneous mixture of the PE throughout the concrete. The indicated dimensions correspond to both the length and diameter of a tube of PE. As can be seen in the table, the PE replaced a fraction of the granite, ultimately giving 10\% by weight PE for the PE-B4C-concrete. The combined volume of the granite and PE in the PE-B4C-concrete was equal to the volume of the granite in the reference concrete. A total of 0.76 wt\% B$_{4}$C was added to the concrete, and it replaced some of the sand as both have similar grain size and mass density. The grain size of the B$_{4}$C was smaller than 1 mm. A picture of the produced concrete is shown Fig. 4, where the B$_{4}$C is not visible but the white spots are the PE beads dispersed in the concrete. The larger black and pink spots are the granite. The total weight fraction of water in the PE-B4C-concrete was around 7\%.

Fig. 5 shows a comparison of the compressive strength of the two concretes after 7 setting days, 28 setting days and 56 setting days. As some of the granite was replaced by PE, it can be expected that the strength of PE-B4C-concrete is lower than that of the standard concrete. 

Fig. 6 shows the mass densities of the concretes at different locations along a single meter long column. As one can see from the figure, PE-B4C-concrete also has a lower density than the standard concrete. This is also due to the replacement of the granite, with a mass density of about 2.6 g/cm$^3$ with the PE, which has a density of 0.95 g/cm$^3$. The lower density of the new concrete thus corresponds to a concrete that is overall 15$\%$ lighter than regular concrete. The figure also illustrates the homogeneity of the mass density of the PE-B4C-concrete throughout the column.

\section{MeV neutron measurements}

In order to evaluate the performance of the concrete in the MeV range, we carried out shielding measurements using the time-of-flight (TOF) technique. The aim of the current measurements was to verify the shielding behavior of the concrete in the MeV neutron energy range. The measurements were carried out using the method of neutron tagging \cite{Scherzinger2015} at the Source-Testing Facility (STF) at the Division of Nuclear Physics at Lund University with neutron energies between 1-7 MeV. Neutron tagging involves measuring the TOF (and thus energy) of individual neutrons emitted from a radioactive source on an event-by-event basis. Ref. \cite{Scherzinger2015} provides a detailed overview of the neutron tagging technique employed at the STF and a further detailed discussion on the facility will be the highlight of a future publication \cite{Hanno}. A general overview is given below.

A basic outline of the experimental method and setup is shown in Fig. 7. For the measurements, a neutron source was located at the center of a water tank with open cylindrical beam ports. A shielding block, of the material of interest, was placed approximately one meter from the position of the source, and a NE-213A detector was located directly behind the block. The radius of the neutron beam on the block ranged from about 12 cm to 14 cm, which depended on the exact source to detector distance which was used for the measurements. Four YAP(Ce) (YAIO$_3$, Ce$^+$ doped) $\gamma$-ray detectors were placed approximately 10 cm from the source and positioned slightly above in order to not interfere with the neutron trajectories. These detectors combined with the NE-213A detector provided the signals for the TOF electronics, which allowed for the determination of the neutron TOF, as described in the following. 

The source employed for the measurements consisted of $^{238}$Pu/$^9$Be. With this source, fast neutrons are released in a two-step process. The actinide $^{238}$Pu first decays by emitting $\alpha$ particles with a mean energy of 5.4891 MeV \cite{nndc}. If the $\alpha$-particle reacts with $^9$Be resulting in a recoiling $^{12}$C, neutrons with energies up to about 11 MeV can be freed \cite{Lorch1973}. The Pu/Be source employed here radiated approximately 2.99$\times$10$^6$ neutrons per second \cite{Source1973}. For further details about the source, see Ref. \cite{ScherzingerPre}.

In the special case that the $^{12}$C nucleus recoils in its first-excited state, the corresponding neutron will be accompanied by a prompt 4.44 MeV de-excitation $\gamma$-ray that is radiated almost isotropically. The resulting radiation field of interest when
this occurs is thus a combination of 4.44 MeV $\gamma$-rays and their associated neutrons. Detection of the 4.44 MeV $\gamma$-ray by a YAP(Ce) detector provided the delayed stop signal for the TOF TDC. Detection of the corresponding neutron in the NE-213A detector, refereed to as a tagged-neutron event, provided the start signal for the TOF TDC. Together, these two signals allowed for the determination of the neutron TOF. The detection of two $\gamma$-rays from a single decay in the source, one in each the NE-213A detector and a YAP(Ce) detector, provided a $\gamma$-flash signal which was used to calibrate the
TOF TDC for tagged neutrons. The time at which the radioactive decay occurred (time zero) was also calculated from this signal using the speed of light and distance to the NE-213A detector position, as described in Ref. \cite{JuliusThesis}.
Note that the tagging technique restricts the maximum available tagged-neutron energies to about 7 MeV due to the energy lost to the 4.44 MeV $\gamma$-ray.  Due to the very clean nature of the TOF data, no pulse-shape discrimination was required, as the events which fell within the TOF range of interest comprised primarily of neutron events \cite{ScherzingerPre}. 

The measured direct TOF spectrum, with no block in the beam, is shown in Fig. 8 in the left panel and the spectrum converted to energy in the right panel. Multiple scattering effects with the water walls of the tank were not included in this conversion. For the shielding measurements, four materials were used as benchmarks. These included blocks of steel, copper and PE in addition to the reference concrete. For the steel, copper, and PE measurements, the total material thickness corresponded to 20 cm for each block. The width and height of each block was 20 cm. For both concrete samples, the total thickness for a single block was 25.5 cm with a width and height of 25.5 cm. Each sample was placed individually in the neutron beam at the shielding block position indicated in Fig. 7. 

A model of the above described setup was implemented in Geant4. This included the direct beam, as shown in Fig. 8, the divergence of the beam provided by the tank, the distances between the source and detectors, the thicknesses of the blocks, and the detector efficiency, as modeled with the Stanton code \cite{Stanton}. A lower energy cutoff of 300 keVee was used during the analysis, which resulted in a lower energy neutron cutoff of 1 MeV for the neutron detection efficiency. The physics list QGSP$\_{}$BERT$\_{}$HP was also used for these simulations and the materials compositions for the concretes are those given in Table 1. The Geant4 simulations were carried out within the ESS Detector Group framework \cite{Kittelmann2014}, which contains developments and tools relevant for neutron shielding calculations. For the comparison of the measurements to simulations, we used directly the TOF instead of the conversion to energy space, as the conversion would require a detailed knowledge of the multiple-scattering effects within the shielding blocks.  

The results with the four benchmark materials are shown in Fig. 9. The points represent the Geant4 simulations and the lines represent the results of the measurements, as described using the methods above. Overall it is seen that there is excellent agreement between the Geant4 simulations and the measurements. This provides confidence that the simulation method and approximations used are sufficient for predicting the shielding performance of the blocks for neutron shielding.

Fig. 10 presents a comparison of the measured reference concrete data, the PE-B4C-concrete data, and the simulations, albeit at a slightly different measurement location than those shown in Fig. 9. Again excellent agreement between the simulations and measurements can be seen. The simulations suggest that over the indicated energy range, there is roughly 40$\%$ fewer neutrons entering the detector for the PE-B4C-concrete relative to the reference concrete.

\section{Conclusions}
In summary, we have developed a specialized concrete for enhanced suppression of neutrons below the few MeV energy range. The concrete was created by adding PE beads and
B$_{4}$C to the material mixture. We found that while the concrete was not as strong as the reference standard concrete, it exhibited a lower density and a significant improvement in
the shielding of neutrons below a few MeVs compared to the standard concrete. An analysis of our measurement data revealed that the new concrete yields around 40$\%$ fewer neutrons,
compared to a standard concrete, in the MeV energy range while Fig. 2 suggests that the improvements are roughly a factor of 10 in the eV-keV range and even better at lower energies.
It is anticipated that the new concrete could find application in either bulk shielding at spallation neutron sources and/or in
specific beamline components such as beamline shielding or instrument caves. It should also be mentioned that the concrete could find applications at reactor- and/or other accelerator-based neutron research
facilities. \\

\section{Acknowledgments}
\noindent This project has received funding from the European Union’s Horizon 2020 research and innovation programme under grant agreement No 654000.

\section*{References}
\bibliographystyle{elsarticle-num}
\bibliography{pbc_paper}

\begin{thebibliography}{10}
\expandafter\ifx\csname url\endcsname\relax
  \def\url#1{\texttt{#1}}\fi
\expandafter\ifx\csname urlprefix\endcsname\relax\def\urlprefix{URL }\fi
\expandafter\ifx\csname href\endcsname\relax
  \def\href#1#2{#2} \def\path#1{#1}\fi

\bibitem{ESS}
S.~Peggs, et~al., ESS Technical Design Report, ESS-2013-001, Lund, Sweden,
  2013.

\bibitem{Cherkashyna2015japan}
N.~Cherkashyna, et~al., Proceedings of ICANS XXI (International Collaboration
  on Advanced Neutron Sources) JAEA-Conf 2015-002 (2015) 479.

\bibitem{Bauer2001}
G.~Bauer, Nucl. Instr. and Meth. in Phys. Res. A 463 (2001) 505.

\bibitem{Moreh2006}
R.~Moreh, et~al., Nucl. Instr. and Meth. in Phys. Res. A 562 (2006) 401.

\bibitem{SINQ}
{SINQ - The Swiss Spallation Neutron Source}, https://www.psi.ch/sinq/.

\bibitem{Wagner2001}
W.~Wagner, {SAFERIB Workshop (CERN, Geneva, Switzerland, 2001), p. 15}.

\bibitem{PSI}
{The Paul Scherrer Institute}, https://www.psi.ch.

\bibitem{Nunomiya2001}
T.~Nunomiya, et~al., Nucl. Instr. and Meth. in Phys. Res. B 179 (2001) 89.

\bibitem{Nakao2004}
N.~Nakao, et~al., Nucl. Instr. and Meth. in Phys. Res. A 530 (2004) 379.

\bibitem{ISIS}
{ISIS - ISIS Home Page}, http://www.isis.stfc.ac.uk/.

\bibitem{RAL}
{Rutherford Appleton Laboratory - Science and Technology Facilities Council},
  http://www.stfc.ac.uk/about-us/where-we-work/rutherford-appleton-laboratory/.

\bibitem{SNS}
{Spallation Neutron Source}, https://neutrons.ornl.gov/sns.

\bibitem{DiJulio2016sns}
D.D.DiJulio, et~al., J. Phys. Conf. Ser. 746 (2016) 012033.

\bibitem{Oakridge1958}
E.P.Blizard, J.~M.Miller, Radiation Attenuation Characteristics Of Structural
  Concrete, Oak Ridge National Laboratory, ORNL-2193, 1958. Available online at
  http://web.ornl.gov/info/reports/1958/3445603503297.pdf.

\bibitem{Oakridge1970}
F.~A.~R. Schmidt, The Attenuation Properties Of Concrete For Shielding Of
  Neutrons Of Energy Less Than 15 MeV, Oak Ridge National Laboratory,
  ORNL-RSIC-26, 1970. Available online at
  http://web.ornl.gov/info/reports/1970/3445603230340.pdf.

\bibitem{Kratochvil2014}
J.~Kratochv\'{i}l, et~al., Adv. Mater. Res. 1000 (2014) 16.

\bibitem{Palomino2008}
L.~A. {Rodr{\'{i}}guez Palomino}, et~al., JINST 3 (2008) P06005.

\bibitem{Calzada2011}
E.~Calzada, et~al., Nucl. Instr. and Meth. Phys. Res. A 651 (2011) 77.

\bibitem{Park2014}
S.J.Park, et~al., J. Nucl. Materials 452 (2014) 205.

\bibitem{GEANT4}
Geant4, http://www.geant4.org.

\bibitem{Geant4Ref1}
S.~Agostinelli, et~al., Nucl. Instr. Meth. A 506 (2003) 250.

\bibitem{DiJulio2016ess}
D.D.DiJulio, et~al., J. Phys. Conf. Ser. 746 (2016) 012032.

\bibitem{Geant4physics}
{Geant4 Reference Physics Lists}, http://geant4.cern.ch/support/
  proc$\_$mod$\_$catalog/physics$\_$lists/referencepl.shtml.

\bibitem{Bert63}
H.~Bertini, Phys. Rev. 131 (1963) 1801.

\bibitem{Bert69}
H.~Bertini, Phys. Rev. 188 (1969) 1711.

\bibitem{Oakridge1960}
W.~R. Burrus, Radiation transmission through boral and similar heterogeneous
  materials consisting of randomly distributed absorbing chunks, Oak Ridge
  National Laboratory, ORNL-258, 1960. Available online at
  http://web.ornl.gov/info/reports/1960/3445603612641.pdf.

\bibitem{Sweden1960}
F.~{\AA}kerhielm, Transmission of Thermal Neutrons through boral, Aktiebolaget
  Atomenergi, Stockholm, Sweden, AE-24, 1960. Available online at
  http://www.iaea.org/inis/collection/NCLCollectionStore/$\_$Public/38/088/38088366.pdf.

\bibitem{DTI}
{Danish Technological Institute}, http://www.dti.dk/.

\bibitem{DTIreport}
T.~Svensson, C.~Pade, Neutron Shielding Concrete - Development of mix design
  and documentation of selected properties, Danish Technological Institute,
  version 2, 2016.

\bibitem{Scherzinger2015}
J.~Scherzinger, et~al., Appl. Rad. and Isotoptes 98 (2015) 74.

\bibitem{Hanno}
H.~Perrey, Manuscript in Preparation.

\bibitem{nndc}
http://www.nndc.bnl.gov/nudat2/.

\bibitem{Lorch1973}
E.~Lorch, J. Appl. Radiat. Isot. 24 (1973) 585.

\bibitem{Source1973}
Exactly 4.26$\times$10$^{6}$ neutrons per second. {Calibration certified at The
  Radiochemical Centre, Amersham, England HP7 9LL on 3 September, 1973}.

\bibitem{ScherzingerPre}
J.~Scherzinger, et~al., Manuscript submitted to Appl. Rad. and Isotoptes (2016)
  Preprint available https://arxiv.org/abs/1611.00213.

\bibitem{JuliusThesis}
J.~Scherzinger, Neutron Irradiation Techniques, PhD Thesis, Lund University,
  2016.

\bibitem{Stanton}
N.~R. Stanton, Monte Carlo program for calculating neutron detection
  efficiencies in plastic scintillator, Ohio State Univ. Research Foundation,
  Columbus (USA), C00-1545-921971. Available online at:
  https://www.osti.gov/scitech/biblio/7179467.

\bibitem{Kittelmann2014}
T.~Kittelmann, et~al., J. Phys. Conf. Ser. 513 (2014) 022017.

\end{thebibliography}

\begin{table}[h]
\caption{Material properties (weight percent and density) for the two concretes used in the simulations.}
\begin{center}
\begin{tabular}{|c|c|c|}
\hline
  & Reference Concrete  & PE-B4C-concrete    \\
\hline
O   & 51.04\%   & 46.06\% \\
Ca  & 7.08\%    & 8.05\% \\
Si  & 32.50\%   & 28.4\% \\
Al  & 3.68\%    & 2.34\% \\
Fe  & 1.15\%    & 0.837\%  \\
Mg  & 0.235\%   & 0.195\%  \\
Na  & 1.05\%    & 0.613\%  \\
K   & 2.11\%    & 1.25\% \\
S   & 0.235\%   & 0.276\% \\
Cl  & 0.00301\% & 0.00353\% \\
H   & 0.782\%   & 2.362\%\\
Ti  & 0.0903\%  & 0.0517\%\\
P   & 0.04520\% & 0.0259\%\\
C   &           & 8.93\%\\
B   &           & 0.596\% \\
\hline
Density & 2.34 (g/cm$^{3}$) & 1.97 (g/cm$^{3}$) \\ 
\hline
\end{tabular}
\end{center}
\label{default}
\end{table}%

\cleardoublepage
\begin{table}[h]
  \caption{The ingredients used to create 1 m$^3$ of each of the concretes \cite{DTIreport}}
  \begin{tabular}{|l|l|l|l|l|}
    \hline
    \multirow{2}{*}{Materials} &
      \multicolumn{2}{c|}{Reference Concrete}  &
      \multicolumn{2}{c|}{PE-B4C-concrete}  \\
    & kg & Vol. \% & kg & Vol. \% \\
    \hline
    Cement       & 350.0  & 11.7 & 350.0  & 11.7\\
    \hline
    Water        & 155.3  & 15.5 & 155.3  &  15.5\\
    \hline
    Admixture    & 0.75   & 0.1  & 0.75 &  0.1\\ 
    \hline
    Sand         & 794.8  & 30.3 & 776.4 & 29.6\\
    \hline
    Granite 4/8  & 258.8  & 9.9  &     &   \\
    \hline
    Granite 8/16 & 794.8  & 30.3 & 508.3 & 19.4\\
    \hline
    PE 2.5 mm    &        &     & 100.7 & 10.4\\
    \hline
    PE 5.0 mm    &        &     & 102.2 & 10.5\\
    \hline
    B$_{4}$C     &        &     & 15.1 &  0.6\\
    \hline    
  \end{tabular}
\end{table}

\cleardoublepage

\begin{figure}[!t]
\centering
\includegraphics[width=140mm]{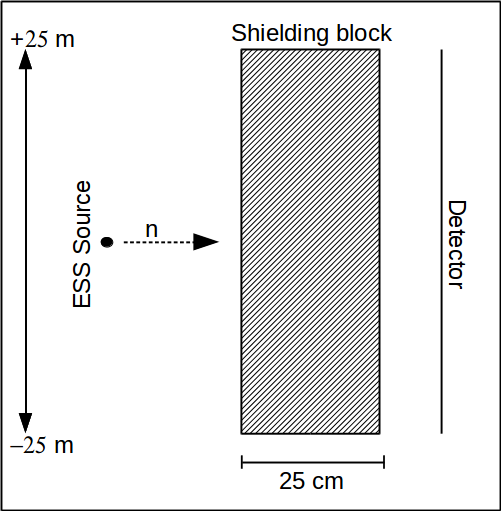}
\caption{A sketch of the geometry used for the Geant4 simulations for the results shown in Figs. 2 and 3. The ESS source is incident on various shielding materials of 25 cm in thickness and the neutron detector is placed behind
the material position. The dimensions extend to $\pm$25 m in the lateral plane.}
\label{fig:flowchart}
\end{figure}

\begin{figure}[!t]
\centering
\includegraphics[width=140mm]{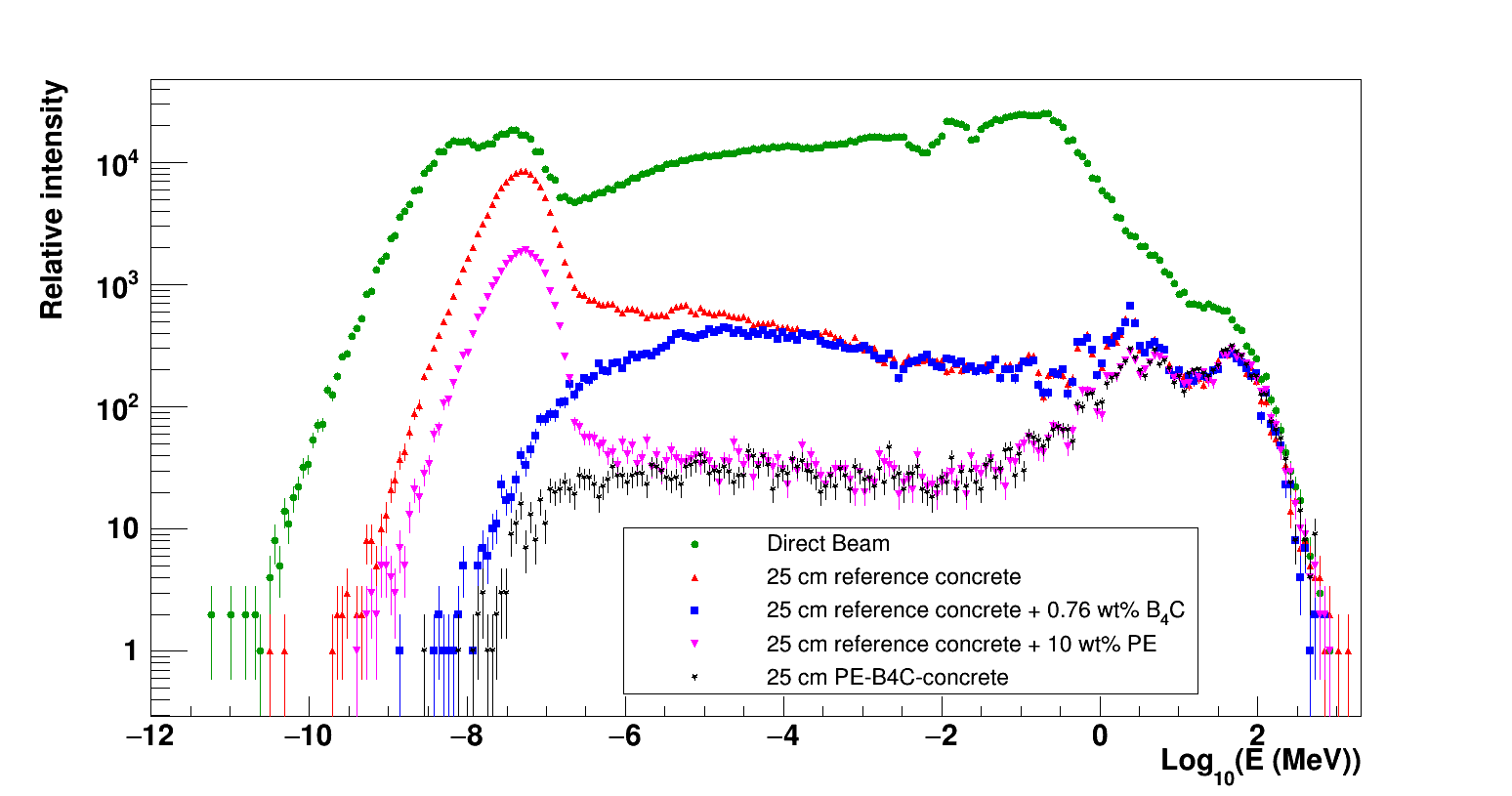}
\caption{Penetration neutron energy spectra for various concrete compositions calculated using Geant4 as described in the text.
The green curve is the incident neutron beam and the remaining curves show the energy spectra for the indicated material compositions. The results were calculated with the Geant4 geometry shown in Fig. 1.}
\label{fig:flowchart}
\end{figure}

\begin{figure}[!t]
\centering
\includegraphics[width=140mm]{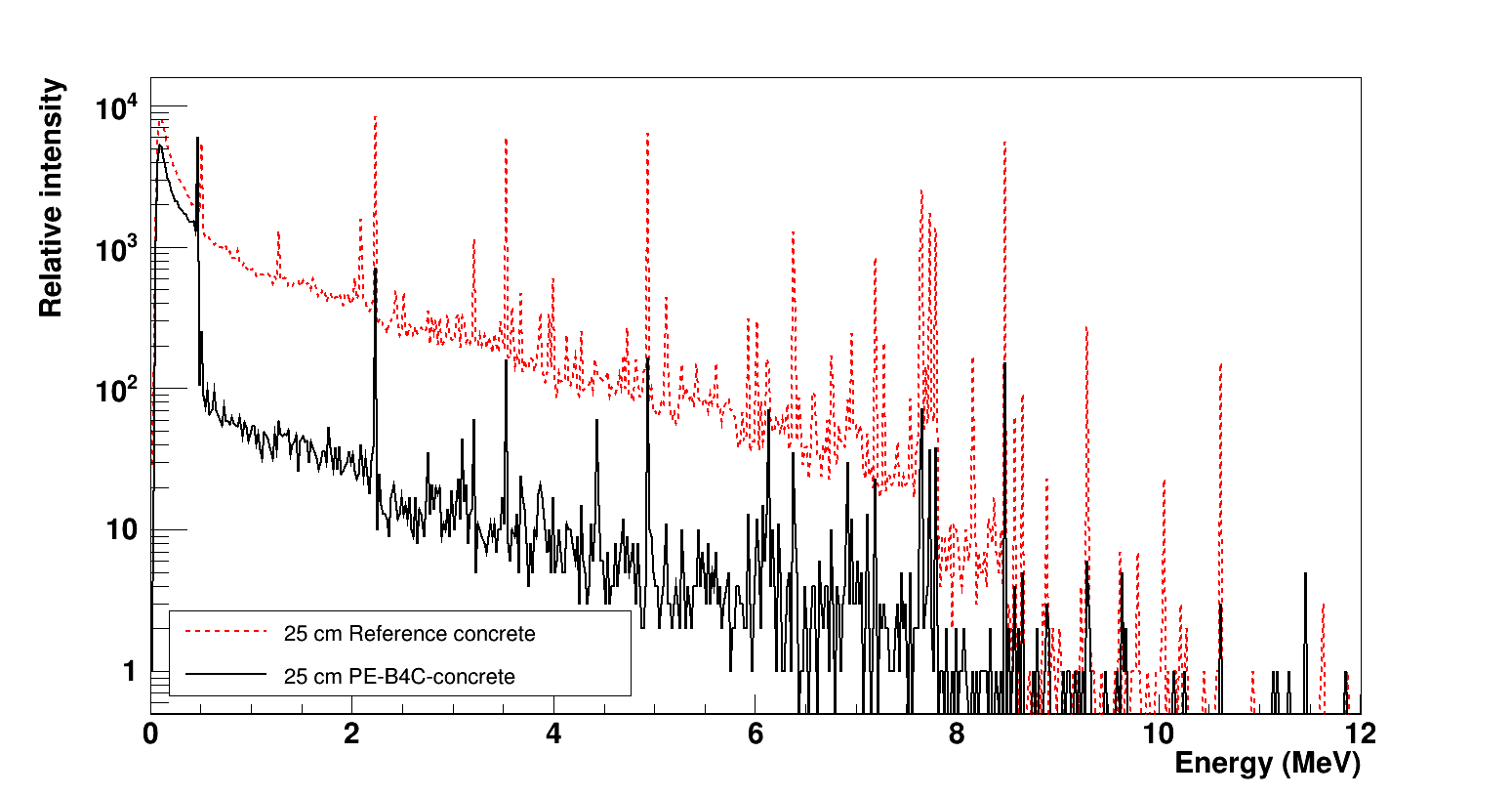}
\caption{Geant4 predictions of the prompt-photon emission of the reference concrete and the PE-B4C-concrete. The results were calculated with the Geant4 geometry shown in Fig. 1.}
\label{fig:flowchart}
\end{figure}

\begin{figure}[!t]
\centering
\includegraphics[width=140mm]{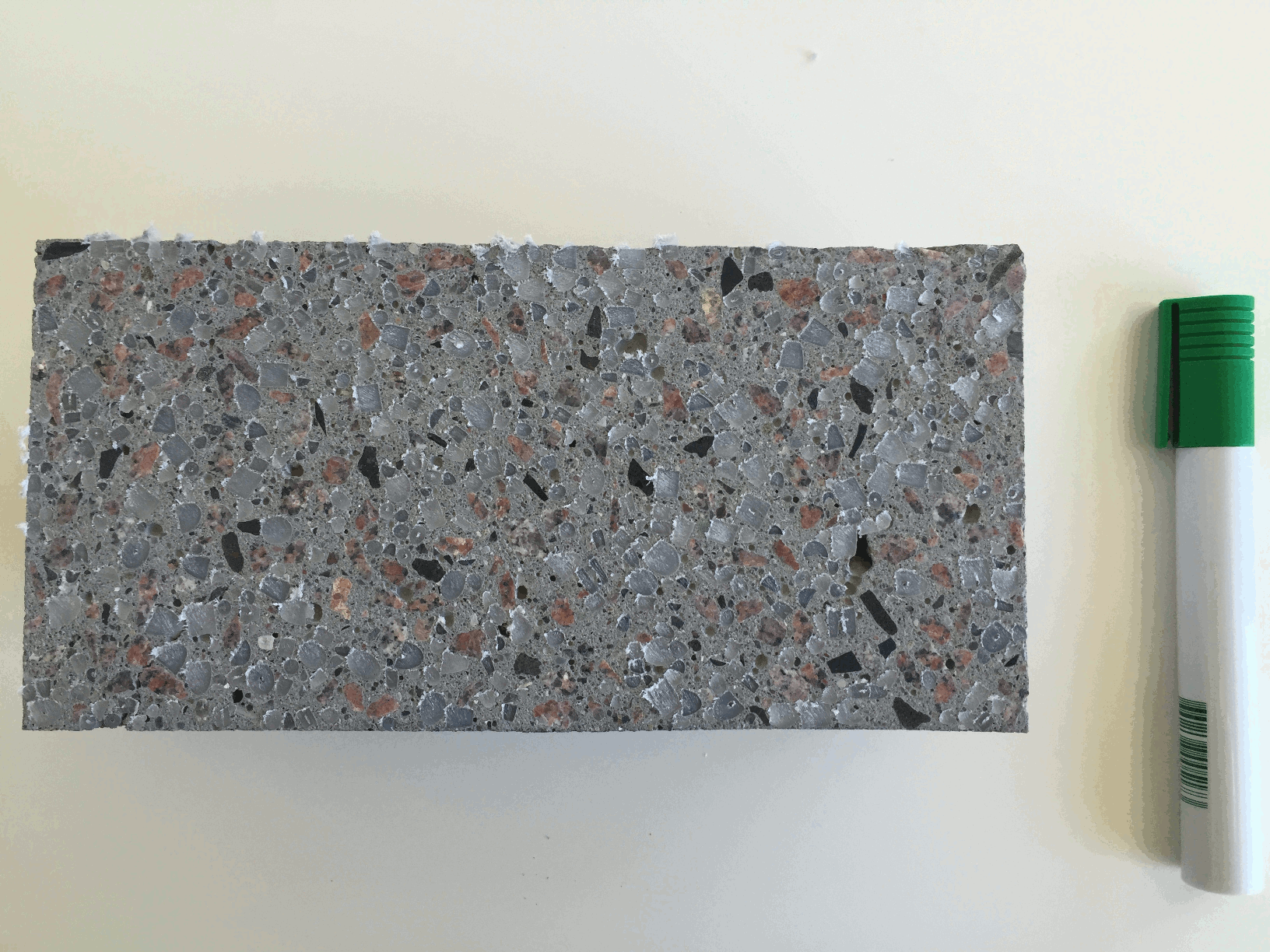}
\caption{Photo of a cross-section of PE-B4C-concrete, the marker is shown for reference.}
\label{fig:flowchart}
\end{figure}

\begin{figure}[!t]
\centering
\includegraphics[width=140mm]{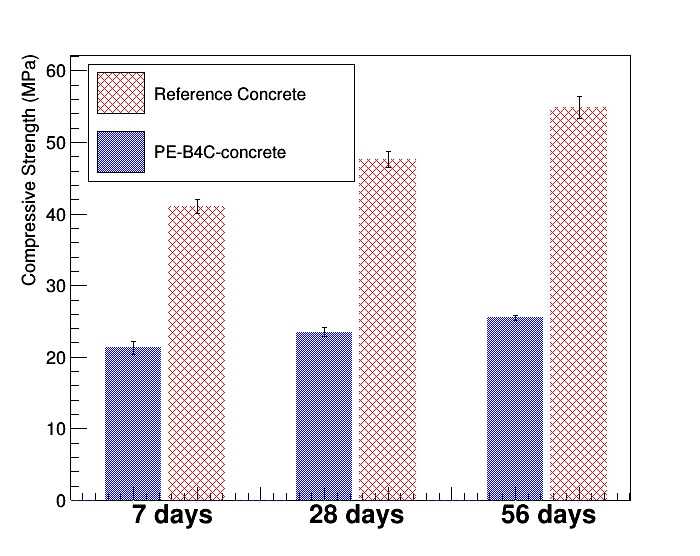}
\caption{The compressive strength of the reference concrete and PE-B4C-concrete after 7, 28 and 56 setting days \cite{DTIreport}.}
\label{fig:flowchart}
\end{figure}

\begin{figure}[!t]
\centering
\includegraphics[width=140mm]{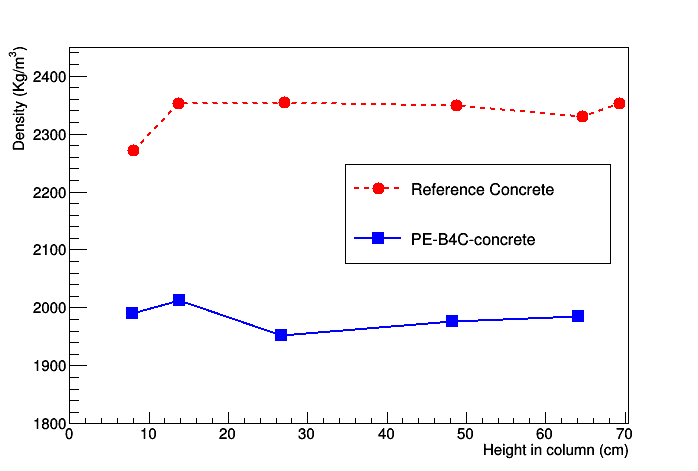}
\caption{The mass density of the reference concrete and PE-B4C-concrete at different locations throughout a column of the indicated concrete \cite{DTIreport}.}
\label{fig:flowchart}
\end{figure}

\begin{figure}[!t]
\centering
\includegraphics[width=140mm]{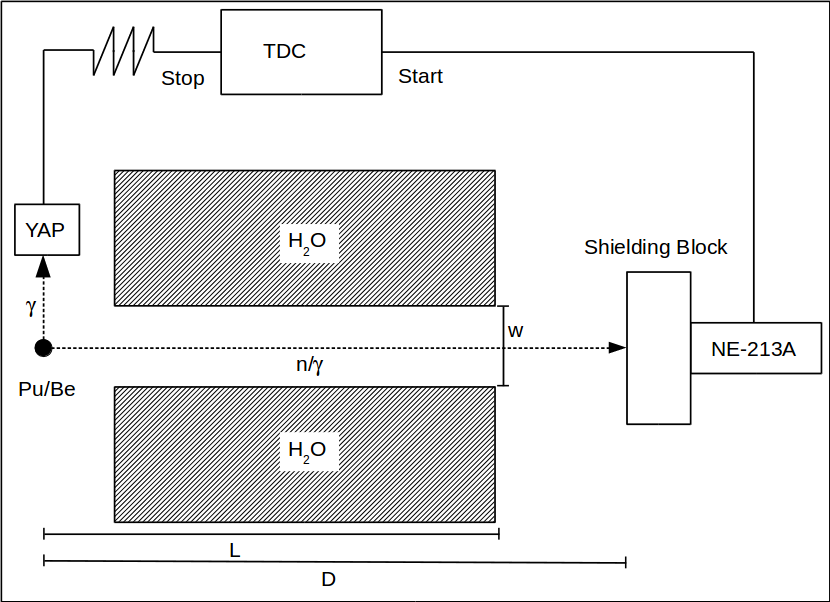}
\caption{A simplified block diagram of the experimental neutron tagging method and setup. The Pu/Be source was located at the center of a water tank, represented by the hashed sections. Only a section of the tank
  is shown for simplification. Four YAP(Ce) detectors were located slightly above the Pu/Be source and the shielding blocks were placed a distance $D$ from the source with a NE-213A detector directly behind.
  The distance $D$ was around 1 meter, the distance $L$=70 cm was the collimation length provided by the tank and $w$=17 cm was the width of the port opening. Neutrons and photons emitted from the Pu/Be source and detected by the NE-213A detector provided the start signal for the TDC while the detection of a coincident photon by one of the four YAP(Ce) detectors provided a delayed stop signal for the TDC. See the text for additional details.}
\label{fig:flowchart}
\end{figure}

\begin{figure}[!t]
\centering
\includegraphics[width=140mm]{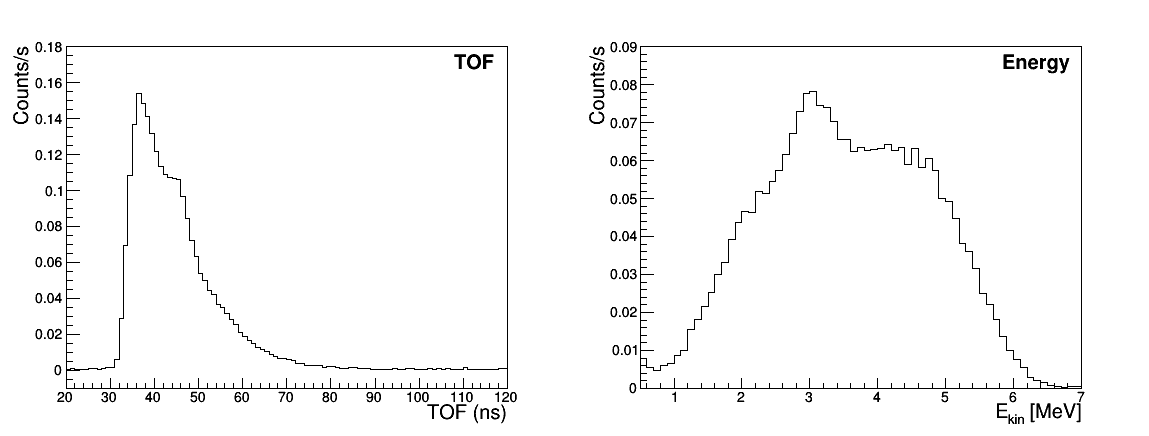}
\caption{The left panel shows the detected neutron TOF spectrum without any shielding blocks in the beam path. The right panel shows the spectrum converted to energy space, as described in the text.}
\label{fig:flowchart}
\end{figure}

\begin{figure}[!t]
\centering
\includegraphics[width=140mm]{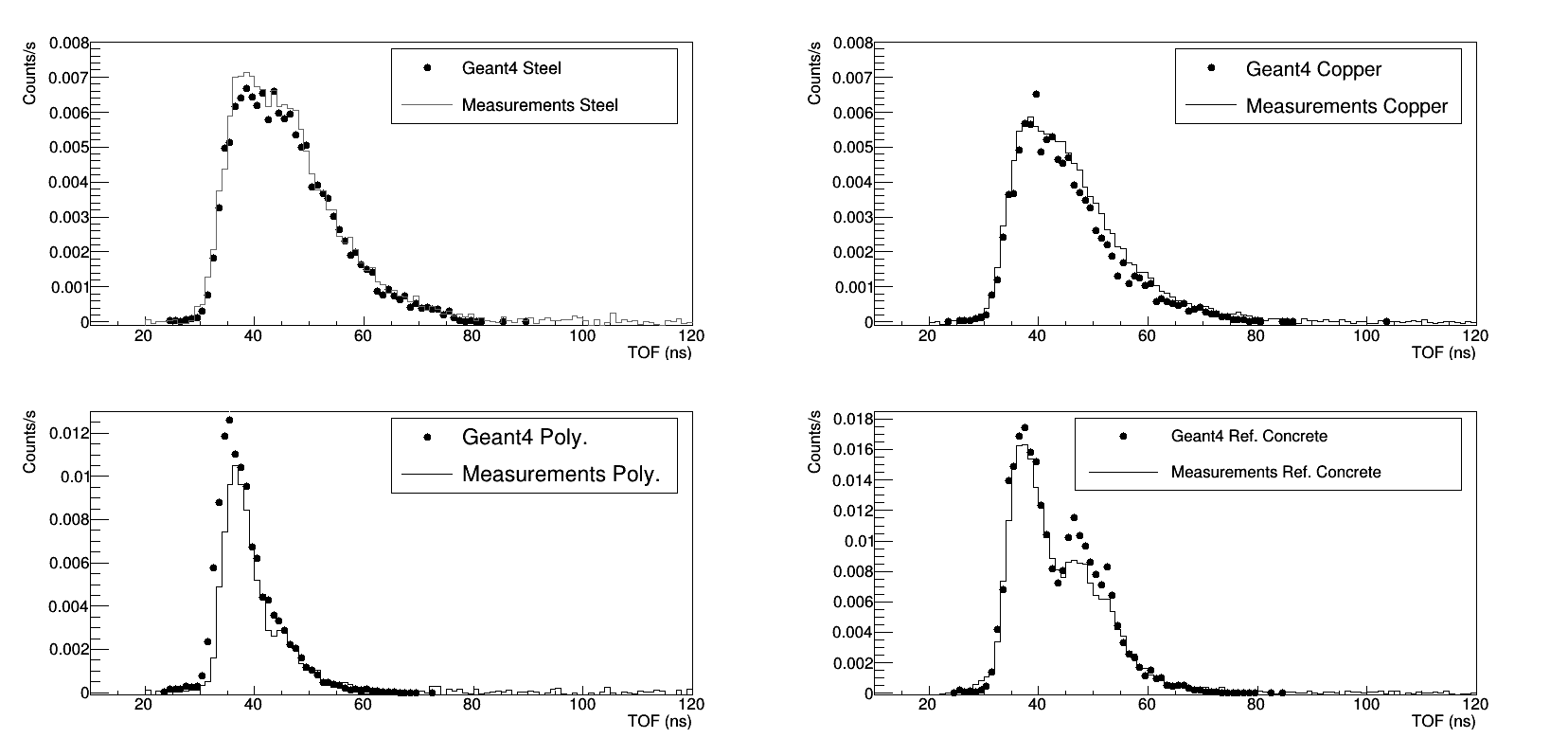}
\caption{A comparison of measurements and simulations for the four benchmark materials. }
\label{fig:flowchart}
\end{figure}

\begin{figure}[!t]
\centering
\includegraphics[width=140mm]{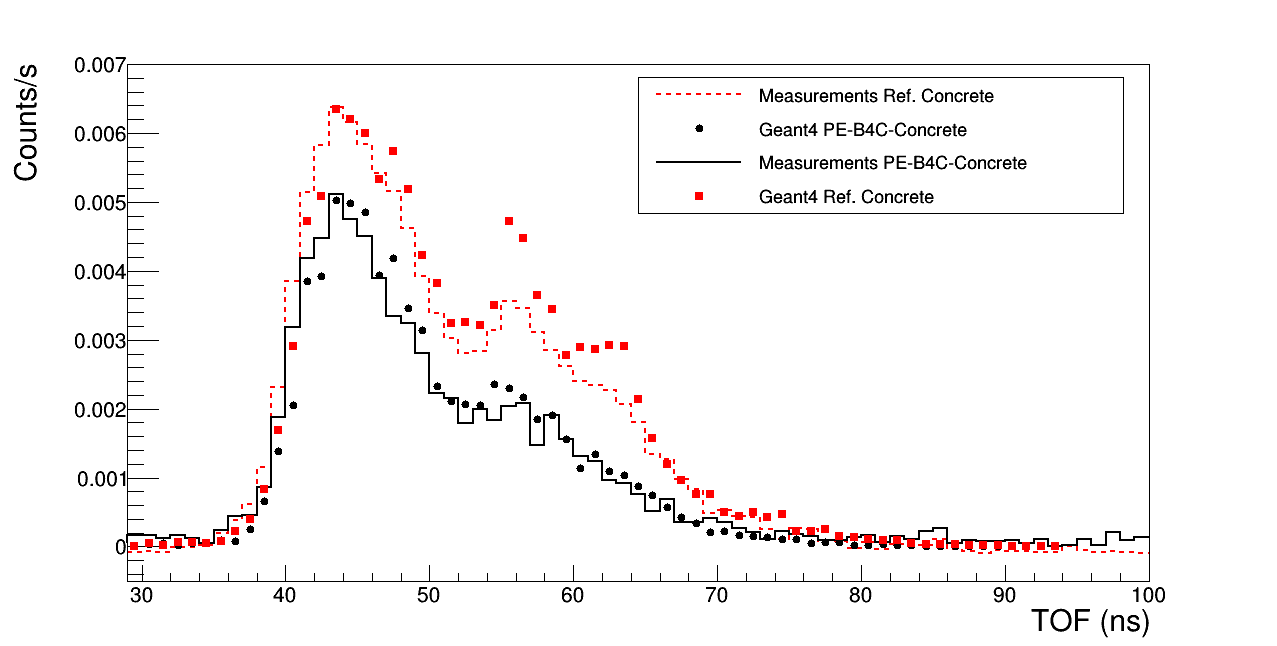}
\caption{A comparison of measurements and simulations for the reference concrete and PE-B4C-concrete.}
\label{fig:flowchart}
\end{figure}

\end{document}